\documentclass{kapproc} % Computer Modern font calls
\newcommand{\eqref}[1]{Eq.~(\ref{#1})}
\usepackage{amssymb}
\usepackage{procps}
\usepackage[dvips]{graphicx}
\upperandlowercase
\kluwerbib
\begin{document}

\articletitle{Gravitational wave interactions with magnetized plasmas}

\articlesubtitle{}

\author{Joachim Moortgat \& Jan Kuijpers}

\affil{Department of Astrophysics, Radboud University Nijmegen\\
PO Box 9010, 6500 GL Nijmegen, The Netherlands}
\email{moortgat@astro.ru.nl, kuijpers@astro.ru.nl}

\begin{abstract}
Gravitational waves (GWs) propagating through a uniformly magnetized plasma interact directly with the magnetic field and excite magnetohydrodynamic (MHD) waves with both electromagnetic and matter components. We study this process for arbitrary geometry in the MHD approximation and find that all three fundamental MHD modes -- slow and fast magnetosonic, and Alfv\'en -- are excited depending on both the polarization of the GW and the orientation of the ambient magnetic field. The latter two modes can interact coherently with the GW resulting in damping of the GW and linear growth of the plasma waves.
\end{abstract}


\section{Introduction}\label{sec::intro}
This decade is expected to witness the historical first direct detection of gravitational waves with detectors such as (Advanced) LIGO, VIRGO, TAMA and others. 
Gravitational waves are emitted by highly energetic events that occur at relatively large distances. Because the GW amplitude falls off with distance from its source, the signal that reaches Earth is exceedingly weak and can only be filtered from a noise signal with some theoretical knowledge of the expected waveforms. To identify a detection of a GW burst, any additional electromagnetic signature of such an event would be extremely useful.

It so happens that many of the proposed GW sources are embedded in a strong magnetic field. Examples are rapidly spinning neutron stars with a small oblateness that precess, accrete, or have an $r$-mode instability, supernovae core collapse and bounce, newly born `boiling' and oscillating neutron stars and magnetars (high frequency GW: \cite{kostasfreq}), magnetars with crust fracturing (low frequency GW: \cite{papadopfreq}) and coalescing compact binaries in which at least one component is a magnetic neutron star.

In the last case, maximum GW luminosities of the order of $10^{55}$~erg/s (\cite{janka}) are released into a wound-up magnetic field of field strength up to $10^{12}$ -- $10^{15}$~Gauss (\cite{ibrahim}) . 
We investigate whether these extreme space-time distortions 
perturb the ambient magnetic field sufficiently to produce an observable electromagnetic counterpart of the GW burst.

In these proceedings we focus our attention on a discussion of the relevant physics and leave the mathematical details to \cite{moortgatII}.
Gaussian geometrized units are used throughout this discussion ($G=c=1$) and Latin indices are used for time-space components ($a = 0 \ldots 3$).

\section{Coupling of the GW to the magnetic field}
The interaction of dynamical space-time with matter and energy is described by Einstein's field equations (EFE). For gravitational waves interacting with a magnetofluid the EFE can be linearized in the GW amplitude $h$ and linear perturbations in the energy and momentum density $\delta T^{ab}$:
%---------------------------------------------%
\begin{eqnarray}\label{eq::efe}
G^{ab} \simeq -\frac{1}{2} \Box h^{ab} &=& 8 \pi \delta T^{ab}.
\end{eqnarray}
%---------------------------------------------%
In the transverse-traceless gauge, a GW propagating in the $z$ direction only has two independent components $h_+ (z,t)$ and $h_\times (z,t)$, corresponding to different polarizations. The only components of $\delta T^{ab}$, in the rest frame of a perfect magnetofluid,  that couple to the GW and can not be removed by a gauge transformation depend on the magnetic field.
Explicitly, \eqref{eq::efe} reduces to:
%.....................................................%
\begin{equation}\label{eq::gwevolution}
\Box h_+  (z,t) =  4 B_x^0 \delta B_x (z,t) , \quad
 \Box h_\times (z,t)  = 4 B_x^0 \delta B_y (z,t),
\end{equation}
%.....................................................%
where the ambient magnetic field is chosen to lie in the $x$-$z$ plane: $\vec{B} = \vec{B}^0 + \delta \vec{B}$ and $\vec{B}^0 = B^0 (\sin\theta, 0, \cos\theta)$.
\eqref{eq::gwevolution} are evolution equations for the GW. 

Similarly, 
we have derived evolution equations for the magnetic field by solving a closed set of magnetohydrodynamic (MHD) equations in an unspecified GW metric. Together with \eqref{eq::gwevolution} these can be solved to find a self-consistent dispersion relation for the coupled gravitational-plasma waves (\cite{moortgatII}). However, we can approximate the GW as a driving wave propagating at the speed of light in the limit  
\begin{equation}\label{eq::limit}
\frac{8\pi}{\mu_0}  (B^0_x)^2 < \omega \Delta k, \quad (\Delta k = k- \omega).
\end{equation}

\section{Alfv\'en, slow and fast magneto-acoustic waves}
In the approximation of a GW driver, the evolution equations for the magnetic field can be solved and look like:
%.....................................................%
\begin{equation}\label{eq::bx}
\delta B_x \propto \frac{1}{2} h_+ B_x^0, 
\qquad 
\delta B_y \propto \frac{1}{2} h_\times B_x^0.
\end{equation}
%.....................................................%
These results are reminiscent of the spatial deviations of test masses in interferometers such as LIGO [$\delta x = \frac{1}{2} (h_+ x_0 + h_\times y_0)$ and
$\delta y = \frac{1}{2} (h_\times x_0 - h_+ y_0)$].

The solution for $\delta B_x$ corresponds to a compressional fast magnetosonic wave (MSW) with both electromagnetic and gas properties. Coherent interaction with the GW is possible when the phase velocity of the MSW approaches that of the GW. In a Poynting flux dominated plasma where the Alfv\'en velocity $u_\mathrm{A}$ is relativistic and much larger than the sound velocity, this limit is satisfied and the perturbations are allowed to grow linearly with distance:
%.....................................................% 
\begin{eqnarray}
\delta B_x (z,t) &\simeq& \frac{h_+}{2} B^0 \sin\theta\ \omega z\  \Im\left[\mathrm{e}^{i\omega (z-t)} \right].
\end{eqnarray}
%.....................................................% 
The phase velocity of the slow MSW is always much smaller than the fast mode, so it can never interact coherently with the GW.

The second expression in \eqref{eq::bx} corresponds to non-compressional shear Alfv\'en waves. The condition for coherent interaction with the GW is more stringent because its phase velocity $u_{\mathrm{A}\|} = u_\mathrm{A} \cos\theta$ has to approach the velocity of light, but at the same time its amplitude is $\propto B^0_x \propto u_\mathrm{A} \sin\theta$. Therefore, in the case of coherent interaction the amplitude of the Alfv\'en waves is suppressed by a small factor $\theta \ll 1$.

%.....................................................%
\begin{equation}\label{eq::Alfvengrowth}
\delta B_y (z,t) \simeq \frac{h_\times}{2}B^0 \theta \  \omega z\ \Im[\mathrm{e}^{i \omega (z-t)}] + {\mathcal O}[\theta^2].
\end{equation}
%.....................................................%

As was mentioned in the previous section, the GW only interacts directly with the magnetic field, and in particular the plasma motion in a GW is generally non-compressional. However, in a perfectly conducting plasma the particles are `glued' to the magnetic field lines and the electromagnetic Maxwell equations couple to the matter conservation laws through the current density. 
Consequently, pressure, density and magnetic field gradients, currents and a drift velocity are also excited in the MSW, whereas the non-compressional Alfv\'en waves cause a divergence of the electric field and a corresponding charge density fluctuation.

\section{GWs propagating through a relativistic wind}
In many of the GW sources mentioned in Sect~\ref{sec::intro}, the spinning matter winds up the magnetic field and causes a collimated relativistic outflow of charged particles in a magnetized plasma wind or jet. This wind is already present before the cataclysmic event such as a binary merger. Therefore the wind has had time to expand over large distances before it is overtaken by the GW. To study the interaction between the GW and the wind we can simply Lorentz boost the results of the previous sections to the frame of an observer looking at the relativistic outflow. The exact expressions for all the wave components can be found in \cite{moortgatII}. The general result is that the plasma wave amplitudes are typically suppressed by a factor $\Gamma^{-2}$, where $\Gamma$ is the Lorentz factor of the wind. This suppression is due to the red-shifted GW frequency and magnetic field in the wind frame. However, this is compensated by the large interaction length scale in the extended wind.% allows for interaction over very large lengthscales.
%, which more than makes up for the $1/\Gamma^2$ suppression.

\section{Conclusions}
We have found that GWs propagating through a magnetized plasma excite all three fundamental MHD modes. Furthermore, if the plasma is magnetically dominated, coherent interaction is possible with the Alfv{\'e}n and fast MSW, allowing these waves to grow. The total energy transferred from the GWs to the plasma is proportional to the square of the ambient magnetic field, the interaction length scale and the GW frequency and amplitude. The most favourable astrophysical sources are therefore merging (magnetic) neutron star binaries and young vibrating magnetars, that have the strongest known magnetic fields and emit GWs with the highest frequencies.

However, even in the most extreme GW sources, it is not yet clear whether this will in fact produce an observable electromagnetic signature directly related to the GW emission (for some numerical estimates see \cite{moortgatI}).
At present, we are investigating whether nonlinear effects in a dilute plasma can lead to radio emission which can be detected by LOFAR.

{
%\chapbblname{moortgat_nikhef}
%\bibliography{prd}
}

\end{document}